\documentclass[10pt]{iopart}

\usepackage{iopams}  

\usepackage{epsfig}
\usepackage{dcolumn}
\usepackage{bm}

\eqnobysec

\newcommand{\be}{{\bf e}}
\newcommand{\bx}{{\bf x}}
\newcommand{\by}{{\bf y}}

\newcommand{\bp}{{\bf p}}
\newcommand{\bq}{{\bf q}}

\begin{document}

\jl{1}

\title{Field theory approach to quantum interference in chaotic
  systems}

\author{Jan M\"uller and Alexander Altland}

\address{Institut f\"ur Theoretische Physik, Z\"ulpicher Str.~77,
  50937 K\"oln, Germany} 

\date{\today}

\begin{abstract}
  We consider the spectral correlations of clean globally hyperbolic
  (chaotic) quantum systems. Field theoretical methods are applied to
  compute quantum corrections to the leading (`diagonal') contribution
  to the spectral form factor. Far--reaching structural parallels, as
  well as a number of differences, to recent semiclassical approaches
  to the problem are discussed.
\end{abstract}

\pacs{03.65.Sq, 03.65.Yz, 05.45.Mt}



\section{Introduction}
Except for a few prominent
counterexamples~\cite{a-Keat91,a-BoGeGiSc91,a-ZaDuDe95}, the low
energy physics of practically all chaotic quantum systems is governed
by the universal spectral correlations of Wigner and Dyson's random
matrix (RM) ensembles~\cite{a-BoGiSc84}. Yet in spite of its ubiquity,
and notwithstanding a number of significant recent
advances~\cite{a-SiRi01,a-Si02,a-Muel03,a-TuRi03,a-Speh03,a-HeMuBrHa_04,a-MuHeBrHaAl04},
the correspondence above is not yet fully understood theoretically.
Specifically, the `non--perturbative' aspects of the problem --- which
manifest themselves, e.g., in the low energy profile of spectral
correlations --- are not under quantitative control.  Some time ago,
the introduction of a field theoretical approach, similar in spirit to
the $\sigma$--models of disordered fermion systems, added a new
perspective to the problem~\cite{a-MuKh95,a-AnSiAgAl96}. This
so--called `ballistic $\sigma$--model' describes chaotic systems in
terms of a field theory in classical phase space.  Remarkably, it
provides a faithful description of RM spectral correlations already on
the most elementary mean field level where fluctuations inhomogeneous
in phase space are neglected; `all' that remains to prove universality
is to show that these inhomogeneities indeed become inessential in the
long time limit --- an expectation backed up by the long time
ergodicity of chaotic systems.

Unfortunately, however, this latter task soon proved to be
excruciatingly difficult. In this paper we shall concentrate on the
perhaps most serious of these problems, the seeming incapability of
the new approach to correctly describe even the lowest order quantum
interference corrections (`weak localization corrections' in the
jargon of mesoscopic physics) to physical observables: In a
semiclassical manner of speaking, `quantum interference' is a process
wherein two initially identical --- modulo the notorious uncertainty
introduced by the non--vanishing of Planck's constant --- Feynman
trajectories split up and later recombine to an overall phase coherent
structure (see figure \ref{sr_pair}). This mechanism is at the root of
practically all quantum phenomena distinguishing disordered or chaotic
quantum systems from their classical limits. It is closely tied to the
notion of the Ehrenfest time --- the time it takes for the separation
of two trajectories to grow from Planck scales to macroscopic scales.
Irritatingly, however, the field theory formalism appeared to be
incapable of describing the initial $\hbar$--uncertainty triggering
these phenomena. Deferring for a more detailed discussion to
section~\ref{sec:field-theor-form} below, let us try to outline the
essence of the problem: loosely speaking, the field degrees of freedom
of the ballistic $\sigma$--model describe the joint propagation of
retarded and advanced Feynman amplitudes along classical trajectories
in phase space. Previous works effectively did not allow for
deviations between the two amplitudes. At this level of approximation,
the retarded and the advanced reference path are strictly identical
and the $\hbar$--quantum uncertainty essential to initiate the
formation of quantum interference corrections is absent.  Equally
important, points in phase space belonging to \emph{different\/}
classical trajectories remain uncorrelated. This implies that the
theory will not be able to describe the relaxation into the uniform
mean field configuration (i.e.~will not be able to predict RMT
behaviour.)

A phenomenological solution to this problem was proposed by Aleiner
and Larkin~\cite{a-AlLa96,a-AlLa97,a-TiLa04}. Building on the insight
gained in previous work, they added a diffusive contribution
(formally, a second order elliptic operator) to the action of the
model. Multiplied by a coupling constant of ${\cal O}(\hbar)$, this
term introduced a sufficient amount of `fuzziness' to the problem to
initiate quantum interference processes.  Although the extra
contribution to the action could not be derived from first principles,
AL argued that it ought to be present on physical grounds (viz.~to
mimic the diffractive aspects of the propagation of quantum states.)

It is the purpose of this paper to demonstrate that, in fact, no
diffraction terms are needed to describe quantum interference within
the framework of the ballistic $\sigma$--model.  Our analysis will
hinge on an aspect of the theory that has been noticed
before~\cite{a-EfScTa04} yet did not receive sufficient attention: the
$\sigma$--model is not a \emph{local\/} field theory in phase space;
by construction, and in accord with the principles of the uncertainty
relation its maximal resolution is limited to Planck cells of
extension $\sim \hbar^f$, where $f$ denotes the number of degrees of
freedom.  We will show that this non--locality suffices to describe
quantum interference in far--reaching analogy with recent
semiclassical approaches~\cite{a-SiRi01,a-MuHeBrHaAl04} to the
problem.\footnote[1]{It is due to mention, though, that our analysis,
  too, necessitates the ad--hoc addition of an extra contribution to
  the action of the native model.  Yet, in a sense to be qualified
  below, this term serves purely regulatory purposes.  Coupled to the
  theory at a strength parametrically weaker than that of the AL term,
  it does not affect the dynamics at times $t\lesssim t_{\rm E}$.} In
recent work~\cite{a-TiKaLa04}, similar ideas have been
applied to compute (in a non--field theoretical setting) weak
localization corrections of a quantum map (viz. the standard map or
kicked rotor).

Specifically, we will consider the spectral two--point correlation
function $R_2(\omega)$ at energies $\omega$ larger than the single
particle spacing $\Delta$. We will show that the expansion of $R_2$ in
the small parameter $s^{-1}\equiv (\pi \omega/\Delta)^{-1}$ agrees
with the prediction of RMT. (In a manner that largely parallels our
present analysis, the same result has recently been obtained by
periodic orbit theory~\cite{a-MuHeBrHaAl04}.) The extensibility of the
analysis to the perturbatively inaccessible regime $s<1$ remains an
open issue.

The rest of the paper is organized as follows: To facilitate the
comparison with the field theoretical formalism, we begin by reviewing
some of the recent developments in the semiclassical approach to
quantum chaos (section~\ref{sec:semicl-backgr}). In
section~\ref{sec:field-theor-form} we turn to the field theoretical
approach and apply it to the perturbative expansion of the two--point
correlation function. We conclude in section~\ref{sec:discussion}.

\section{Semiclassical Background}\label{sec:semicl-backgr}

We are interested in the behaviour of globally hyperbolic (chaotic)
quantum systems at time scales $t$ larger than the ergodic time
$t_{\rm erg}$\footnote[1]{Formally, $t_{\rm erg}$ is defined as the
  inverse of the first non--vanishing Perron--Frobenius eigenmode. In
  fact, all our results can be generalized to general mixing rather
  than just uniformly hyperbolic systems. The point is that mixing
  implies ergodicity and non--integrability, and hence any mixing
  system will appear to have a constant global Lyapunov exponent when
  evaluated on time scales $t>t_{\rm erg}$.} yet smaller than the
Heisenberg time $t_{\rm H}\equiv 2\pi\hbar/ \Delta$.  (The first
condition implies that non--universal aspects of the classical
dynamics are inessential, the second that concepts of perturbation
theory (in the parameter $\tau\equiv t/t_{\rm H}$) are applicable.)

To describe correlations in the spectrum of the system, we 
consider the  two--point correlation function
\begin{equation}\label{r2}
  R_2(\omega) \equiv \Delta^2 \langle \rho(E+\omega/2)
  \rho(E-\omega/2)\rangle_E -1
\end{equation}
and its Fourier transform
\begin{equation}\label{formfactor}
  K(t) \equiv \frac{1}{\Delta}\int\rmd\omega\,
  \rme^{-\frac{\rmi}{\hbar}\omega t} R_2(\omega),
\end{equation}
the spectral form factor. Here, $\rho(E)$ is the energy dependent
density of states (DoS) and $\langle \dots \rangle_E$ denotes
averaging over a sufficiently large portion of the spectrum centered
around some reference energy $E_0$.

In semiclassics, the spectral form factor is expressed as 
\begin{equation*}
K_{\rm sc}(\tau)=\Bigl\langle\sum_{\gamma\gamma'}A_\gamma
  A_{\gamma'}^* {\rm e}^{\rmi(S_\gamma-S_{\gamma'})/\hbar}
  \delta\Big(\tau-\textstyle{\frac{T_\gamma+ T_{\gamma'}}{2t_{\rm H}}}
  \Big)\Bigr\rangle,
\end{equation*}
where $\sum_{\gamma\gamma'}$ is a double sum over periodic orbits
$\gamma$ and $\gamma'$, $S_\gamma$ the classical action of the orbit,
$T_\gamma$ its revolution time, and $A_\gamma$ its classical stability
amplitude.

Before turning to a more detailed discussion, let us briefly summarize
the main results recently obtained for the semiclassical form
factor: For times $\tau<1$, $K_{\rm sc}$ can be expanded in a series
in $\tau$. As shown by Berry~\cite{a-Berr85}, the dominant
contribution to this expansion $K^{(1)}_{\rm sc}= 2\tau$, is provided
by pairs of identical $(\gamma=\gamma')$ or mutually time reversed
($\gamma={\cal T} \gamma'$) paths. (Throughout we focus on the case of
time reversal and spin rotation invariant systems --- orthogonal
symmetry.)

\begin{figure}[h]
  \centering
  \epsfig{file=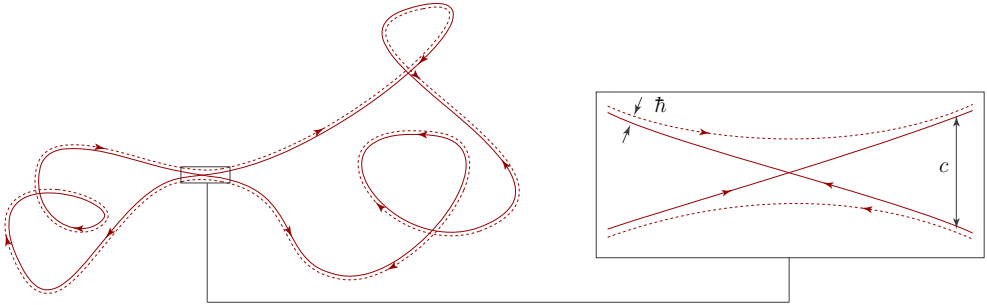, width=\textwidth}
  \caption{\label{sr_pair}
    Cartoon of a pair of topologically distinct paths,
    $(\gamma,\gamma')$ contributing to the first quantum correction to
    the spectral form factor.  Notice that $\gamma$ and $\gamma'$
    differ in exactly one intersection point (crossing vs.~avoided
    crossing). Inset: blow--up of the intersection region.}
\end{figure}

All corrections to the leading contribution $K^{(1)}$ hinge on the
mechanism of quantum interference alluded to in the introduction.
E.g., the sub--dominant contribution, $K_{\rm sc}^{(2)}$, to the form
factor is provided by pairs $(\gamma,\gamma')$ that are nearly
identical except for one `encounter region':\footnote{Notice that a
  path of duration $t\gg t_{\rm erg}$ generally contains many self
  intersections in \emph{configuration space\/}.} In this region one of
the path self--intersects while its partner just so avoids the
intersection (cf.~figure \ref{sr_pair}).  (Alternatively, one may think
of two trajectories that start out nearly identical, then split up and
later recombine to form an interfering Feynman amplitude pair.)  The
two paths are, thus, topologically distinct yet may carry almost
identical classical action~\cite{a-AlLa96}. Specifically, Sieber and
Richter~\cite{a-SiRi01} have shown that for sufficiently shallow self
intersections (crossing angle in configuration space of ${\cal
  O}(\hbar)$) the action difference $|S_\gamma - S_{\gamma'}|\lesssim
\hbar$. For these angles, the duration of the encounter process is of
the order of the Ehrenfest time $t_{\rm
  E}=\lambda^{-1}\ln(c^2/\hbar)$, where $\lambda$ is the phase space
average of the dominant Lyapunov exponent of the system and $c$ a
classical reference scale (see below) whose detailed value is of
secondary importance. This identifies $t_{\rm E}$ as the minimal time
required to form quantum corrections to the form factor (as well as to
other physical observables~\cite{a-AlLa96}).  Throughout we shall
assume $t_{\rm erg}<t_{\rm E} < t < t_{\rm H}$, where the condition
$t_{\rm erg}< t_{\rm E}$ is imposed to guarantee that for time scales
$t>t_{\rm E}$, the system indeed behaves universally.  (For $t_{\rm
  erg}>t_{\rm E}$, the time window $t_{\rm E}<t<t_{\rm erg}$ is
characterized by the prevalence of correlations that are
non--universal yet quantum mechanical in nature.)
\begin{figure}[h]
  \centering
\epsfig{file=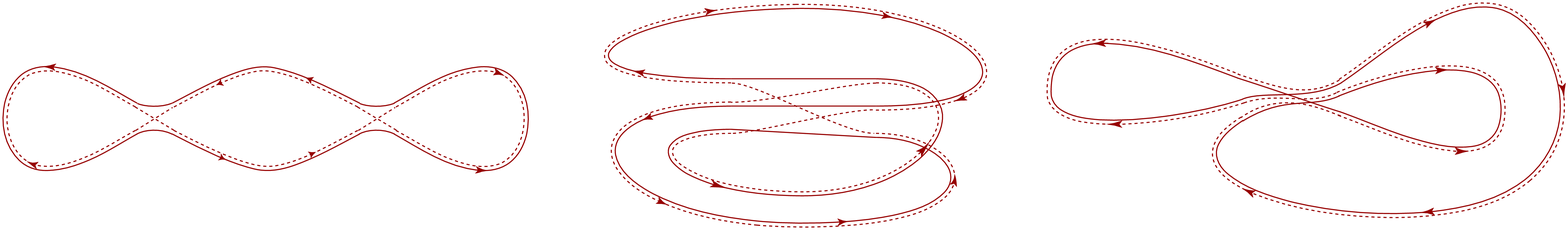, width=\textwidth}
  \caption{\label{tau_3_terms}
    Cartoon of three classes of orbit pairs that contribute to the
    expansion of the form factor at order $\tau^3$. (The
    triple--encounter region shown in the two figures on the right is
    the analog of the Hikami hexagon familiar from the impurity
    diagram approach to disordered systems.) The existence of the
    middle pair does not rely on time reversal invariance.}
\end{figure}

Summation over all Sieber--Richter pairs~\cite{a-SiRi01} leads to the
universal result $K_{\rm sc}\simeq K^{(1)}_{\rm sc}+K^{(2)}_{\rm
  sc}=2\tau - 2\tau^2$, which is consistent with the short time
expansion of the random matrix form factor
\begin{equation}\label{krm}
    K_{\rm RM}(\tau)\stackrel{0\leq \tau\leq 1}{=} 2\tau - \tau \ln(1+2\tau).
\end{equation}
At higher orders in the $\tau$--expansion, orbit pairs of more complex
topology enter the stage. (For some families of pairs contributing to
the next leading correction, $K^{(3)}$, see figure \ref{tau_3_terms}.)
The summation over all these pairs~\cite{a-MuHeBrHaAl04} --- feasible
under the presumed condition $t_{\rm erg}<\tau$ --- obtains an
infinite $\tau$--series which equals the series expansion of the RMT
result (\ref{krm}).\footnote{However, as is indicated by the notorious
  non--analyticity of $K_{\rm RM}(\tau)$ at $\tau=1$~\cite{b-Meht91},
  the form factor at times $\tau\ge 1$ appears to be beyond the reach
  of semiclassical summation schemes.} It is also noteworthy that both
the topology of the contributing orbit pairs and the combinatorial
aspects of the summation are in one--to--one correspondence to the
impurity--diagram expansion~\cite{a-SmLeAl98} of the spectral
correlation function of disordered quantum systems.

\begin{figure}[h]
  \centering
  \epsfig{file=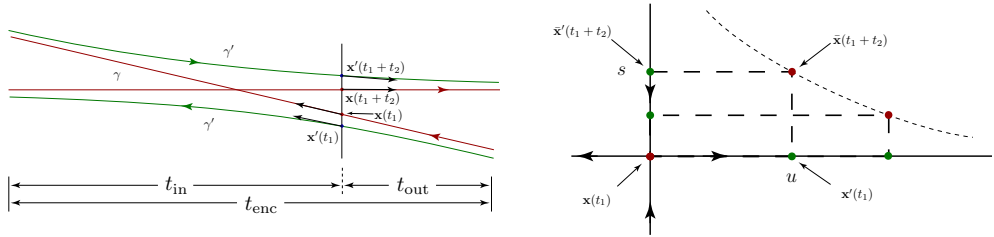, width=\textwidth}
  \caption{\label{tr_encounter}
    The structure of the encounter region. The picture on the right
    shows how the parallelogram spanned by the four points evolves in
    time $t_1$, while its symplectic area $us$ is conserved.}
\end{figure}

Central to our comparison of semiclassics and field theory below will
be the understanding of the encounter regions where formerly pairwise
aligned orbit stretches reorganize. The analysis of these objects is
greatly facilitated by switching from the configuration space
representation originally used in~\cite{a-SiRi01} to one in phase
space~\cite{a-Muel03,a-TuRi03,a-Speh03}. In the following we briefly
discuss the phase space structure of the regions where periodic orbits
meet.  In section~\ref{sec:pert-expans} we will compare these
structures to the --- somewhat different --- field theoretical variant
of encounter processes.

Considering the correction $K_{\rm sc}^{(2)}$ as an example, we note
that the encounter region contains four orbit stretches in close
proximity to each other (cf.~figures \ref{sr_pair},
\ref{tr_encounter}): two segments $\bx(t_1)$ and $\bx'(t_1)$ of the
orbits $\gamma$ and $\gamma'$ traversing the encounter region and the
\emph{time reversed\/}\footnote{In a standard position--momentum
  representation $\bx=(\bq,\bp)$, time reversal is defined as
  $\bar{\bx}\equiv(\bq,-\bp)$.} $\bx(t_1+t_2)$ and $\bx'(t_1+t_2)$ of
the trajectories reentering after one of the loops adjacent to the
encounter region has been traversed ($t_2$ is the duration of the loop
traversal and $t_1$ parameterizes the time during which the encounter
region is passed).  To describe the dynamics of these trajectory
segments, it is convenient to introduce a Poincar\'e surface of
section ${\cal S}$ transverse to the trajectory $\bx(t_1)$.  For a
system with two degrees of freedom (a billiard, say), ${\cal S}$ is a
two--dimensional plane slicing through the three--dimensional subspace
of constant energy in phase space. We chose the origin of ${\cal S}$
such that it coincides with $\bx(t_1)$.  Introducing coordinate
vectors ${\bf e}_u$ and ${\bf e}_s$ along the stable and unstable
direction in ${\cal S}$, the three points $\bar{\bx}(t_1+t_2)$,
$\bx'(t_1)$ and $\bar{\bx}'(t_1+t_2)$ are then represented by the
coordinate pairs $(u,s), (u,0)$ and $(0,s)$, respectively. (Notice
that the trajectory $\bx'$/$\bar{\bx}'$ traverses the encounter region
on the unstable ($s=0$)/stable $(u=0)$ manifold thus deviating
from/approaching the reference orbit $\bx$.)

The above coordinate system is optimally adjusted to a description of
the two main characteristics of the encounter region: its duration
$t_{\rm enc}$ and the action difference $S_\gamma-S_{\gamma'}$.
Indeed, it is straightforward to show that the total action difference
is simply given by the area of the parallelogram spanned by the four
reference points in phase space, $S_\gamma-S_{\gamma'} =
us$~\cite{a-Speh03}. As for the encounter duration, let us assume that
the distance between the orbit points may grow up to a value $c$
before they leave what we call the `encounter region'. (It is natural
to identify $c$ with the typical phase space scale up to which the
dynamics can be linearized around $\bx(t_1)$, however, any other
classical scale will be just as good.) After the trajectory $\bx$ has
entered the encounter region, it takes a time $t_{\rm in} \sim
\lambda^{-1} \ln(c/s)$ to reach the surface of section and then a time
$t_{\rm out}\sim \lambda^{-1} \ln(c/u)$ to continue to the end of the
encounter region. (Here, $\lambda$ is the Lyapunov exponent of the
system. Thanks to the assumption $t_{\rm erg} \ll t_{\rm E}$,
$\lambda$ may be assumed to be a `self averaging quantity', constant
in phase space.)  The total duration of the passage is thus given by
$t_{\rm enc}(u,s)\equiv t_{\rm out}+t_{\rm in} \sim \lambda^{-1} \ln
(c^2/(u s))$. The action difference of orbit pairs contributing
significantly to the double sum must be small, $|S_\gamma-
S_{\gamma'}|= us \lesssim \hbar$.  Consequently, $t_{\rm enc} \gtrsim
t_{\rm E}\equiv \lambda^{-1} \ln(c^2/\hbar)$, where $t_{\rm E}$ is the
Ehrenfest time introduced above.  (Notice that both
$S_\gamma-S_{\gamma'}$ and $t_{\rm enc}$ depend only on the product
$us$. While the individual coordinates $u$ and $s$ depend on the
positioning of the surface of section, their product $us$ is a
canonical invariant and, therefore, independent of the choice of
${\cal S}$.)

Having discussed the microscopic structure of the encounter region, we
next need to ask a question of statistical nature: given a long
periodic orbit $\gamma$ of total time $t$, what is the number
$N(u,s,t)d ud s$ of encounter regions with Poincar\'e parameters in
$[u,u+d u]\times [s,s+d s]$? (To each of these encounter regions there
will be exactly one topologically distinct partner orbit $\gamma'$
that is identical to $\gamma$ in all other $(N-1)$ encounters. Thus,
$N(u,s,t)duds$ is the number of Sieber--Richter pairs for a given
parameter configuration and $\int\rmd u\rmd s N(u,s,t)$ is the total number
of Sieber--Richter pairs.) Since the times $t_1$ and $t_2$ defining
the two traversals of the encounter region are arbitrary (except for
the obvious condition $|t_1-t_2|>t_{\rm enc}$), $N$ is proportional to
the double integral $N(u,s,t)d ud s\propto {1\over
  2}\int_{0,|t_2-t_1|>t_{\rm enc}}^td t_1d t_2\, P_{\rm ret}(u,s,t_2)
d ud s$. The integrand, $P_{\rm ret}$ is the probability to propagate
from the point $(0,0)$ in the Poincar\'e section to the time--reverse
of $(u,s)$ in time $t_2$.  Since $t_2>t_{\rm E}>t_{\rm erg}$, this
probability is constant and equals the inverse of the volume
$\Omega=2\pi\hbar t_{\rm H}$ of the energy shell, $P_{\rm
  ret}(u,s,t_2)=\Omega^{-1}$.  Thanks to the constancy of $P_{\rm
  ret}$, the temporal integrals can be performed and we obtain
$N(u,s,t) \propto t(t-2t_{\rm enc})/2\Omega$.  The normalization of
$N$ is fixed by noting that the temporal double integral weighs each
encounter event with a factor $t_{\rm enc}$.  The appropriately
normalized number of encounters thus reads $N(u,s) = {t(t-2t_{\rm
    enc})\over 2t_{\rm enc} \Omega}$. Substitution of $N(u,s,t)$ into
the Gutzwiller sum obtains
\begin{equation*}
\eqalign{K^{(2)}(\tau)&=\sum_\gamma |A_\gamma|^2
  \delta\Bigl(\tau-{t_\gamma\over t_{\rm H}}\Bigr)\int_{-c}^c d ud s\,
  N(u,s,t)2\cos(us/\hbar)\\
  &={\tau^2 \over 2\pi \hbar} \int_{-c}^c d ud
  s\, \Bigl({t\over t_{\rm enc}(u,s)}
  -2\Bigr)\cos(us/\hbar)\stackrel{\hbar\to 0}{=}-2\tau^2,}
\end{equation*}
where we used the sum rule $\sum_\gamma |A_\gamma|^2
\delta(\tau-{t_\gamma\over t_{\rm H}})=\tau$ of Hannay and Ozorio de
Almeida~\cite{a-HaOz84} and noted that in the semiclassical limit the
first term in the integrand does not contribute (due to the singular
dependence of $t_{\rm enc}$ on $\hbar$.)

Before leaving this section, let us discuss one last point related to
the semiclassical approach: the analysis above hinges on the ansatz
made for the classical transition probability $P_t(\bx,\bx')$ between
different points in phase space. Specifically, a na\"ive
interpretation of ergodicity --- $P_t(\bx,\bx')=\Omega^{-1}={\rm
  const.}$ for times $t>t_{\rm erg}$ --- is too crude to obtain a
physically meaningful picture of weak localization.  One rather has to
take into account that the unstable coordinate, $u(t)$, separating two
initially close ($u(0)\ll c$) points $\bx$ and $\bx'$ grows as $u(t)
\sim u(0)\exp(\lambda t)$. For sufficiently small initial separation,
the time it takes before the region of local linearizability is left,
\begin{equation}\label{tehrenfest}
 \case{1}{2} t_{\rm E}(\bx,\bx') \equiv \frac{1}{\lambda} \ln \frac{c}{u(0)},
\end{equation}
may well be larger than $t_{\rm erg}$. This is important because
during the process of exponential divergence, the probability to
propagate from $\bx$ to the time--reverse $\bar\bx'$ is identically
zero. (Simply because the proximity of $\bx$ and $\bx'$ implies that
$\bx$ and $\bar{\bx}'$ are far away from each other.)  Only after the
domain of linearizable dynamics has been left, this quantity becomes
finite and, in fact, constant:
\begin{equation}\label{transition}
 P_t(\bx,\bx')={1\over \Omega}\Theta(t-t_{\rm E}(\bx,\bx')).
\end{equation}
This concludes our brief survey of the semiclassical approach to
quantum coherence. We next turn to the discussion of the field
theoretical formulation and its structural parallels to the formalism
above.

\section{Field Theoretical Formulation}\label{sec:field-theor-form}
\subsection{Definition of the Model}\label{sec:definition-model}

The ballistic $\sigma$--model is defined by a functional integral ${\cal
  Z}(\omega)=\int {\cal D} T\rme^{-S[T]}$ extending over field
configurations $T(\bx)$ in classical phase space. Its action is
given by $S[T]=S_0[T]+S_{\rm reg}[T]$, where
\begin{equation}
  \label{actiont}
  S_0[T]=\frac{\rmi}{4\hbar}\int(d\bx)\,\tr\Bigl(
  \frac{\omega^+}{2}\sigma_3^{\rm ar}T^{-1}\Lambda T+ 
  T^{-1} \Lambda [H,T]\Bigr)
\end{equation}
is the action of the `native' model~\cite{a-AnSiAgAl96} and $S_{\rm
  reg}$ a regulatory contribution to be discussed momentarily. Here,
$\int(d\bx)\equiv (2\pi\hbar)^{-f+1}\int\rmd\bx\,\delta(E_0-H(\bx))$
is the integral over the $(2f-1)$--dimensional shell $\Omega$ of
constant energy $E_0$\footnote[1]{See~\ref{sec:regularization} for
  details on the definition of this integral.}, $T=T(\bx)$,
$\sigma_3^{\rm ar}$ and $\Lambda$ are matrices whose internal
structure will be discussed momentarily, $H$ the classical Hamilton
function of the system, and $\omega$ the scale at which we are probing
the spectrum. (Within the field theoretical approach it is preferable
to work in energy rather than in time space.)  Importantly, all
products appearing in the action (\ref{actiont}) have to be understood
as Moyal products,
\begin{equation*}
(f g)(\bx) \equiv\rme^{\frac{\rmi\hbar}{2}\partial_{\bx_1} I
  \partial_{\bx_2}} f(\bx_1)g(\bx_2)\big|_{\bx_1=\bx_2=\bx},  
\end{equation*}
where the matrix $I$ is defined through $ \bx^T I \bx'\equiv
\bq\cdot\bp'-\bp\cdot\bq'.  $ For later reference we note that the
Moyal product affords the alternative representation
\begin{equation}\label{moyal_nonloc}
  (fg)(\bx)=\int\frac{d\bx_1}{(\pi\hbar)^f}
  \frac{d\bx_2}{(\pi\hbar)^f}\,\rme^{\frac{2\rmi}{\hbar}\bx_1^TI\bx_2}
  f(\bx+\bx_1)g(\bx+\bx_2).
\end{equation}
Equation (\ref{moyal_nonloc}) makes the `non--locality' inherent to the
action of the ballistic $\sigma$--model manifest: all products involve
a coordinate averaging over Planck cells of volume $\sim
\hbar^f$.  As we shall see below, this
non--locality encapsulates essential aspects of the semiclassical
dynamics discussed in the previous section.

The second contribution to the action
\begin{equation}\label{actionreg}
  S_{\rm reg}[T]=\frac{g_{\rm reg}}{8}
  \int(d\bx)\,\tr\bigl(\partial_i T(\bx)\partial_i T^{-1}(\bx)\bigr)
\end{equation}
serves to damp out singular field configurations (Unlike with most
other field theories, the action of the unregularized model, governed
by the generator $[H,\;]$ of unitary quantum dynamics, does not have
the capacity to self--regularize.)  In~\ref{sec:regularization} we
will argue that a coupling constant $g_{\rm reg}\sim \hbar^2$ suffices
to stabilize the theory.  Coupled to the theory at this strength, the
action $S_{\rm reg}$ does not yet influence the dynamics on the
physically relevant times $t_{\rm E}$.
This stands in contrast to the theory of AL where a second order
derivative term (similar in structure to~(\ref{actionreg}) but with
coupling constant $g_{\rm reg}\sim \hbar$) actively governed the
dynamics at times $t\simeq t_{\rm E}$.

In the original references, the ballistic $\sigma$--model was
introduced as a supersymmetric field theory. However, for the purposes
of our present analysis it will be more convenient to employ the
simpler formalism of the replica trick. Within this approach, the
matrices $T(\bx)\in {\rm Sp}(4R)/{\rm Sp}(2R)\times {\rm Sp}(2R)$ act
in a tensor product of $R$--dimensional replica space, a
two--dimensional space distinguishing between advanced and retarded
propagators (ar--space) and a two--dimensional space (tr--space) whose
presence is required to account for the time reversal invariance of
the system~\cite{a-Wegn79}. Here, ${\rm Sp}(2R)$ is the
$2R$--dimensional symplectic group and $\Lambda=\sigma_3^{\rm
  ar}\otimes\mathbf{1}^{\rm tr}\otimes \mathbf{1}_R$, where
$\mathbf{1}_R$ is the $R$--dimensional unit matrix. While the use of
replicas bars us from performing non--perturbative calculations, it
significantly facilitates the comparison to the semiclassical analysis
above.

\subsection{Two--Level Correlation function}
\label{sec:two-level-corr}

Our goal is to compute the two--level correlation function
$R_2(\omega)$. Expressed in terms of  single particle Green
functions $G^\pm=(E\pm\rmi 0-H)^{-1}$, 
\begin{equation*}
R_2(\omega) = {\Delta^2\over 2\pi^2} {\rm Re}\, \left\langle{\rm
    tr}(G^+(E+\omega/2))\, {\rm tr}(G^-(E-\omega/2))
    \right\rangle_{E,{\rm c}},
\end{equation*}
where $\langle\dots \rangle_{E,{\rm c}}$ denotes the (connected)
average over an energy interval much larger than the inverse of the
smallest time scales in the problem (the inverse of the Lyapunov time,
say.) Within the replica formalism, $R_2$ is obtained by a two--fold
differentiation of the partition function w.r.t.~the energy
parameter:\footnote{This follows from the fact that (by construction)
  \[
    Z(\omega_1-\omega_2)= \langle \det [\rmi G^{+}(E+\omega_1)]^R \det
    [\rmi G^{-}(E+\omega_2)]^R\rangle_E.
  \]
  Using that $\ln z=\lim_{R\to 0} (z^R-1)/R$, it is then
  straightforward to verify that
  \[
  \fl\lim_{R\to 0}{1\over R^2}\,{\rm Re}\,\partial_{\omega_1-\omega_2}^2
  Z(\omega_1-\omega_2)=-{\rm Re}\,\langle\tr\,G^+(E+\omega_1)
  \tr\,G^-(E+\omega_2)\rangle_{E,{\rm c}}=
  -2(\pi/\Delta)^2 R_2(\omega_1-\omega_2).
  \]}
\begin{equation*}
R_2(s) =-\case{1}{2}\lim_{R\to 0}{1\over R^2}{\,\rm Re}\,
\partial^2_s Z(s),
\end{equation*}
where the dimensionless variable $s=\pi \omega/\Delta$.  As long as
we restrict ourselves to perturbative operations, i.e.~an expansion of
the two--level correlation function in a series
\begin{equation}\label{r2s}
  R_2(s) \stackrel{s>1}{=}{\rm Re}\,\sum_{n=2}^\infty c_n (\rmi s^+)^{-n},
\end{equation}
the replica limit $R\to 0$ is well--defined. A straight\-forward
Fourier transformation, $K(\tau) =\pi^{-1}\int\rmd s\,\rme^{-2\rmi
  s\tau} R_2(s)$, shows that the coefficients $c_n$ are related to the
coefficients $d_n$ of the spectral form factor $K(\tau)\equiv
\sum_{n=1}^\infty d_n \tau^n$ through
\begin{equation}\label{dandc}
  d_n = -\frac{(-2)^n}{n!} c_{n+1}.
\end{equation}
In fact, however, there are much more far--reaching analogies between
the temporal and the frequency representation of spectral
correlations: at every given order $n$ various topologically distinct
families of orbit/partner orbit pairs (`diagrams') contribute to the
coefficient $d_n$. Likewise, the expansion coefficients $c_n$ obtain
as sums of Wick contractions of the generating functional $Z(\omega)$.
We shall see that there is an exact correspondence between field
theoretical and semiclassical diagrams (both in topological structure
and numerical value) which simply means that the two approaches
describe spectral correlations in terms of the same semiclassical
interference processes.

\subsection{Quadratic Action}
\label{sec:quadratic-action}

We next turn to the actual expansion of the field integral. For this
purpose, we shall employ the so--called `rational parameterization' of
the coset--valued field $T$. This parameterization is defined by
$T=\mathbf{1}+W$, where
\begin{equation}\label{woff}
W=\left(\matrix{&-B^\dagger\\\cr B&}\right)_{\rm ar}
\end{equation}
is a matrix that anti--commutes with the matrix $\Lambda$ introduced
above. Its off--diagonal blocks take values in the Lie algebra of
${\rm Sp}(2R)$, i.e.~they satisfy the constraint $B^\dagger
=B^\tau\equiv (\rmi\sigma_2^{\rm tr}\otimes \mathbf{1}_R) B^T
(\rmi\sigma_2^{\rm tr}\otimes \mathbf{1}_R)^{-1}$. The principal advantage
of the rational parameterization is that the Jacobian associated to
the transformation from the $T$--matrices to the linear space of
$B$--matrices is unity: $\int {\cal D}T=\int {\cal D}B$.

Substituting this representation into the action, we obtain a series
$S[B]=\sum_{n=1}^\infty S^{(2n)}[B]$, where $S^{(2n)}$ is of $2n$--th
order in $B$. Let us begin by discussing the unregularized quadratic
action
\begin{equation*}
S_0^{(2)}[B]=-\frac{\rmi}{2\hbar}\int (d\bx)\,\tr\left[
B^\dagger \left(\omega^+-[H,\;]\right)B\right],
\end{equation*}
where $[H,\;]$ is the generator of quantum time evolution. Very
little can be said about this generator in concrete terms which means
that the action $S_0^{(2)}$ does not qualify  as a `reference point' of a
perturbative expansion scheme. (Indeed, notice that the projection
$|\alpha \rangle \langle \alpha|$ onto an eigenstate $|\alpha\rangle$
of the Hamilton operator $H$ is annihilated by $[H,\;]$. This means
that the quantum evolution operator possesses a large number of
(nearly) unstable `zero modes' whose action is damped only by the
frequency parameter $\omega$.)

Much more is known about the generator $\{H,\,\}$ of classical
dynamics, where
$\{f,g\}(\bx)\equiv\partial_{\bx_1}^{T}I\partial_{\bx_2}
f(\bx_1)g(\bx_2)\big|_{\bx_1=\bx_2=\bx}$ is the Poisson bracket.
Expanding the Moyal commutator,
\begin{equation*}
[H,B](\bx)=\rmi\hbar\{H,B\}(\bx)+
{\cal O}((\hbar\partial_\bx)^3 B(\bx)),
\end{equation*}
we notice that the quantum generator $[H,\;]$ differs from its
classical counterpart $\{H,\;\}$ by the presence of higher order
derivative terms. In~\ref{sec:regularization} it is shown
that the quadratic regulatory action
\begin{equation*}
S_{\rm reg}^{(2)}[B]= g_{\rm reg} \int (d\bx)\,\tr\bigl(\partial_i
B^\dagger(\bx) \partial_i B(\bx)\bigr)
\end{equation*}
suffices to damp out higher derivatives and hence effectively
projects the quadratic action onto its classical limit.  Assuming
regularization in the above sense, our further discussion will be
built on the action
\begin{equation}\label{actioncl}
S_{\rm cl}^{(2)}[B]=\case{1}{2}\int(d\bx)\,\tr\left[
B^\dagger(\bx) ({\cal L}_\omega B)(\bx)\right],
\end{equation}
where ${\cal L}_\omega\equiv -\rmi\omega/\hbar-\{H,\;\}$. Throughout,
the operator $P_{\omega}\equiv {\cal L}_{\omega,{\rm reg}}^{-1}$ will
play an important role. Here, the subscript `reg' indicates that
$\mathcal{L}_\omega$ acts in a space of functions coarse grained over
cells in phase space of `volume' $(\hbar^2/a)^f$, where $a$ is some
classical reference scale of dimensionality `action' whose detailed
value will not be of much concern. Importantly, $P_\omega$ is not
strictly inverse to the bare Liouville operator (i.e.~the Liouville
operator acting in the space of unregularized functions.),
$\mathcal{L}_\omega P_\omega(\bx,\bx')\not=\delta(\bx-\bx')$. Rather,
the time Fourier transform, $P_t(\bx,\bx')=\delta(\bx-\bx(t))$ can
resolve the definite dynamical evolution generated by the Liouville
operator only up to time scales
\begin{equation*}
\tilde{t}_{\rm E} \equiv \lambda^{-1} \ln\frac{c^2a}{\hbar^2} \sim
2t_{\rm E}.
\end{equation*}
Thereafter, the uncertainty in the resolution of the boundary
conditions (the effect of smoothening) renders the dynamics
unpredictable, i.e.
\begin{equation}\label{crossover}
  P_t(\bx,\bx')=\left\{
    \begin{array}{ll}
\delta(\bx-\bx'(t))&,\ t< \tilde t_{\rm E},\cr
\Omega^{-1}&,\ t>\tilde t_{\rm E}.
    \end{array}\right.
\end{equation}
the crossover between the two regimes takes place over time scales
$\sim \max\{\Delta \tilde{t}_{\rm E},t_{\rm erg}\}$, where
$\Delta\tilde{t}_{\rm E} \ll t_{\rm E}$ is the uncertainty in
$\tilde{t}_{\rm E}$ caused by an eventual non--uniformity of the
Lyapunov expansion.\footnote{The results above apply to uniformly
  hyperbolic systems. In the case of non-uniform hyperbolic systems,
  local fluctuations in the Lyapunov expansion rate $\lambda(\bx)$
  need to be taken into account. The logarithmic mismatch
  $y(\bx,t)=\ln(u(t)/u(0))$ between two trajectories starting at $\bx$
  and $\bx+u(0)\be_u$, respectively, grows as $\dot
  y=\lambda(\bx(t))$. ($\be_u$ is the locally most unstable direction
  in phase space.) Due to inhomogeneities in the expansion rate,
  $y(\bx,t)$ is a fluctuating quantity with mean $y(t)$ and a certain
  width $\Delta y(t)$.  Importantly, an upper bound on fluctuations in
  $y$ is imposed by Oseledec's theorem~\cite{a-Osel68} which states
  that the phase space average $\lambda$ of the Lyapunov expansion
  rate equals the long--time expansion rate of individual trajectories
  almost everywhere: $y(\bx,t)/t\to\lambda$ for $t\to\infty$ for
  almost all $\bx$.  Consequently, $\Delta y(t)\sim t^\alpha$ grows at
  a rate $\alpha<1$. (E.g., the model of statistically independent
  Gaussian fluctuations of the local expansion rate employed by
  AL~\cite{a-AlLa96} leads to $\alpha=\frac{1}{2}$.) By definition of
  $\tilde{t}_{\rm E}$, an phase space distribution of initial
  extension $\sim \hbar^{2f}$ has expanded to classical dimensions
  when $y(t)=\lambda\tilde{t}_{\rm E}$.  Defining $\Delta \tilde
  t_{\rm E}$ as the time uncertainty in $\tilde t_{\rm E}$ (due to
  fluctuations in the local expansion rate), we obtain the estimate
  $\Delta \tilde t_{\rm E}\sim \Delta y(\tilde{t}_{\rm E})/\lambda\sim
  t_{\rm E}^\alpha$. This means that $\Delta \tilde{t}_{\rm E}/
  \tilde{t}_{\rm E}\sim t^{\alpha-1}$ vanishes in the semiclassical
  limit. For finite $\hbar$, the effective relaxation rate of the
  system is set by $\max\{\Delta \tilde{t}_{\rm E},t_{\rm erg}\}$.}
(Notice that in previous discussions of the ballistic $\sigma$--model,
the propagator $P$ was mostly identified with the Perron--Frobenius
operator, i.e.~an object that describes relaxation into a uniform
configuration, $P_t(\bx,\bx')\stackrel{t>t_{\rm erg}}{=}{\rm const.}$
over classically short times. However, it is not at all clear how this
behaviour may be reconciled with the indispensable condition that
$P_t(\bx,\bar \bx')\stackrel{t<t_{\rm E}(\bx,\bx')}{=}0$: for
$|\bx-\bx'|={\cal O}(\hbar)$, the propagator must be able to resolve
fine structures in phase space over times $\sim t_{\rm E}$
parametrically larger than the relaxation time of the
Perron--Frobenius operator. In contrast, equation (\ref{crossover}) is
motivated by the structure of the action, and does resolve the
classical phase space dynamics up to the Ehrenfest time. In fact, we
will see that the scale $\tilde t_{\rm E}>t_{\rm E}$ does not
explicitly enter the results of the theory. The reason is that, due to
a conspiracy of quantum non--locality and chaotic instability, the
dynamics becomes effectively irreversible \emph{instantly\/} after the
Ehrenfest time (on time scales of the order of the inverse Lyapunov
exponent).  Thus, at a time where the artificially introduced smearing
would become virulent, the theory has long become
quantum--unpredictable.)

\subsection{Perturbative Expansion}\label{sec:pert-expans}

We now have everything in store to proceed to the perturbative
expansion of the functional integral. The dominant contribution to the
series~(\ref{r2s}) obtains by integration over the quadratic
action:
\begin{equation}\label{r2s2}
\fl\eqalign{R_2^{(2)}(s)&=-\case{1}{2}\lim_{R\to 0}{1\over R^2}
{\rm Re}\,\partial_s^2\int {\cal D}B
\exp(-S_{\rm cl}[B])\\
&=-\case{1}{2}\,{\rm Re}\lim_{R\to 0}{1\over R^2}
\partial_s^2(\det P_\omega)^{2R^2}
={\rm Re}\,\partial_s^2
\ln\det(P^{-1}_\omega)\stackrel{\omega\ll\hbar/t_{\rm erg}}{\simeq}
{\rm Re}\,(\rmi s^+)^{-2}.}
\end{equation}
This result implies (cf.~equations (\ref{r2s},~\ref{dandc})) $d_1=2$
in accord with the semiclassical analysis.\footnote{It is worthwhile
  to notice that the agreement between semiclassics and field theory
  does not pertain to times $t<t_{\rm erg}$: for these times, short
  periodic orbits traversed more than once influence the behaviour of
  the form factor. For reasons that only partly understood, the
  $\sigma$--model fails to correctly count the integer statistical
  weight associated to the repetitive traversal of periodic orbits.
  The essence of the problem~\cite{i-Mirl99} is that the degrees of
  freedom of the $\sigma$--model (the $B$'s) describe the joint
  propagation of amplitudes locally paired in phase space.  However,
  an $n$--fold repetitive process is governed by the local correlation
  of $2n$ Feynman amplitudes.  Perturbative approaches to the problem
  fail to correctly describe these correlations.  Interestingly, a
  non--perturbative evaluation of the functional integral --- feasible
  in the artificial case of periodic orbits with unit monodromy matrix
  --- leads to the correct result (M.~R. Zirnbauer, unpublished).}

To compute higher order terms in the expansion we need to consider the
non--linear contributions $S^{(2n>2)}$ to the action. Substituting the
representation (\ref{woff}) into the action (\ref{actiont}) we obtain
\begin{equation}\label{s2n}
  S^{(2n)}[B]=\case{1}{2}\int(d\bx) 
  \tr\left[(-B^\dagger B)^{n-1}B^\dagger\mathcal{L}_\omega B\right].  
\end{equation}
By elementary power counting, each matrix $B$ scales as (symbolic
notation) $\sim (\mathcal{L}_\omega)^{-\frac{1}{2}}\sim
\omega^{-\frac{1}{2}}\sim s^{-\frac{1}{2}}$.  This implies that each
vertex $S^{(2n)}$ contributes a factor $\sim (B^\dagger B)^{n-1}
B^\dagger \mathcal{L}_\omega B \sim s^{-n+1}$ to the functional
integral.  Specifically, the dominant correction ($\sim s^{-3}$) to
the leading contribution~(\ref{r2s2}) obtains by first order expansion
in the vertex $S^{(4)}$:
\begin{equation}
   R_2^{(3)}(s)=-{\,\rm Re}\lim_{R\to0}\frac{1}{(2R)^2}\partial^2_s \int
         (d\bx)\langle{\,\rm tr}(B^\dagger B B^\dagger \mathcal{L}_\omega
         B)\rangle_B.
     \label{r2_3}
\end{equation}
We emphasize again that all products of $B$--matrices have to be
understood as Moyal products.  To obtain a convenient representation
of the product of more than two of these matrices, we iteratively
apply the prototype formula equation (\ref{moyal_nonloc}).  A
straightforward calculation then yields the general product formula
\begin{equation*}
\fl (A_1\dots A_{2n})(\bx)=
\int\prod_{i=1}^{2n}\frac{d\bx_i}{(\pi\hbar)^f}\,
\rme^{\frac{\rmi}{\hbar}S(\bx_1,\cdots\bx_{2n})}
A_1(\bx+\bx_1)\dots A_{2n}(\bx+\bx_{2n}),
\end{equation*}
where the bilinear form $S(\bx_1,\cdots\bx_{2n})\equiv
2\sum_{i<j}(-)^{i+j}\bx_i^T I \bx_j$.  Below, we will apply this
formula to the fields $B$ of the theory. In~\ref{sec:regularization}
we show that in this case, all energy coordinates $E_i$ get locked.
Here, we assume a coordinate choice $\bx=(E,t,\by)$ where $E$ is an
energy variable, $\tau$ its canonically conjugate time--like variable
(a coordinate parameterizing the Hamiltonian flow through $\bx$) and
$\by$ a $(2f-2)$--dimensional vector of coordinates transverse to the
flow.  Further, fluctuations in the time--like variables $\tau_i$ are
negligible.  Introducing the shorthand notation $\int_{\bx_i}\equiv
(\pi\hbar)^{-f+1}\int\rmd \bx_i\, \delta(E_i-E_0)\delta(\tau_i)$, we
thus have
\begin{equation}\label{moyal2nB}
(B^\dagger B)^{2n}(\bx)=\int_{\bx_1,\dots,\bx_{2n}}
\hspace{-1cm}\rme^{\frac{\rmi}{\hbar}S(\bx_1,\cdots\bx_{2n})}
B^\dagger(\bx+\bx_1)\dots B(\bx+\bx_{2n}).
\end{equation}
Using this representation in (\ref{r2_3}), applying the contraction
rules (\ref{contraction}) discussed in~\ref{sec:perturbation_theory},
and taking the replica limit we obtain
\begin{eqnarray*}
\fl R_2^{(3)}(s)=
{\rm Re}\Bigl(\frac{\Omega}{t_{\rm H}}\Bigr)^2
\partial_s^2 \int(d\bx)\int_{\bx_1,\dots,\bx_4}\hspace{-8mm}
\rme^{\frac{\rmi}{\hbar}S(\bx_1,\dots,\bx_4)}
P_\omega(\overline{\bx+\bx_1},\bx+\bx_3)
    {\cal L}_{\omega,\bx_4} P_\omega(\bx+\bx_4,\overline{\bx+\bx_2}),
\end{eqnarray*}
where the coordinate subscript in ${\cal L}_{\omega,\bx}$ indicates
the argument on which the Liouvillian acts. The physical meaning of
this expression is best revealed by switching to the Fourier conjugate
picture. Inserting the definition~(\ref{formfactor}) of the form
factor, we obtain
\begin{equation}\label{K_2_tau}
\fl\eqalign{K^{(2)}(\tau) =- 2\tau^2&\frac{\Omega^2}{t_{\rm H}}
\int (d\bx) \int_{\bx_1,\dots,\bx_4}\hspace{-8mm}
\rme^{\frac{\rmi}{\hbar}S(\bx_1,\dots,\bx_4)}\times\\
&\times\int_0^t dt' P_{t-t'}(\overline{\bx+\bx_1},\bx+ \bx_3)
{\cal L}_{t',\bx_4} P_{t'}(\bx+\bx_4,\overline{\bx+\bx_2}),}  
\end{equation}
where $\mathcal{L}_t\equiv \partial_t - \{H,\;\}$. equation (\ref{K_2_tau})
makes the analogies (as well as a number of differences) between the
semiclassical and the field theoretical description of quantum
corrections explicit: central to both approaches are two
semi--loops shown schematically in figure \ref{sr_pair}.
In either case, the proximity of these loops is controlled by phase
factors which contain the coordinates of the end points (in a
canonically invariant manner) as their arguments.  However, unlike
with semiclassics, equation (\ref{K_2_tau}) does not relate the unification
of the two semi--loops to specific periodic orbits.  Rather, the two
halves are treated as independent entities, each described in terms of
its own probability factor $P$.  Relatedly, the phase factor
controlling the proximity of the terminal points does not correspond
to the action difference between two orbits.

The result obtained for $K^{(2)}(\tau)$ in equation (\ref{K_2_tau})
critically depends on the behaviour of the propagator $P_t$ at times
$t\sim t_{\rm E}(\bx,\bx')\sim t_{\rm E}$,
cf.~equations (\ref{tehrenfest},~\ref{transition}).  Specifically, we
shall use that $\partial_t P_t(\bar{\bx},\bx')=\Omega^{-1}
\delta(t-t_{\rm E}(\bx,\bx'))$, where $\delta(t)$ is some smeared
$\delta$--function whose detailed functional structure is not of much
importance. (All we shall rely upon is $\int\rmd t \delta(t)
f(t)\simeq f(t)$ for functions that vary slowly on the scales where
$\delta(t)$ varies.) We also note that the Poisson bracket
$\{H,f\}(\bx) \sim \partial_t f(\bx)$ effectively differentiates along
the trajectory through $\bx$.  However, the time $t_{\rm
  E}(\bx,\bx')=t_{\rm E}(\by,\by')$ defined in
equation (\ref{tehrenfest}) depends only on the coordinates transverse
to the trajectory.  This implies
$\{H,P_t(\bar{\bx},\bx')\}=\{H,P_t(t_{\rm E}(\bx,\bx'))\}=0$. We thus
conclude that the action of ${\cal L}_t$ on the function $P$ is given
by ${\cal L}_{\bx,t} P_t(\bar{\bx},\bx')\simeq \Omega^{-1}
\delta(t-t_{\rm E}(\bx,\bx'))\simeq \Omega^{-1} \delta(t-t_E)$. To
understand the meaning of the second approximation, notice that it
takes a time $t_E$ before the bulk of the Planck cell to which the
points $\bx$ and $\bx'$ belong has grown to classical scales $c$. For
times $t>t_E$, the fraction of the Planck cell which has not yet
acquired macroscopic dimensions shrinks exponentially on the classical
Lyapunov scale $\lambda^{-1} \ll t_E$. This means that
$t_E(\bx,\bx')\simeq t_E$ up to an insignificant uncertainty of ${\cal
  O}(\lambda^{-1})$.  Using these results, as well as the
normalization relations $\int(d\bx)=t_{\rm H}$ and
$\int_{\bx_1,\dots,\bx_{2n}}\rme^{\frac{\rmi}{\hbar}S(\bx_1,\dots,\bx_{2n})}=1$,
we obtain
\begin{eqnarray*}
\fl\eqalign{K^{(2)}(\tau) &\simeq -2\tau^2\frac{1}{t_{\rm H}}
  \int (d\bx) \int_{\bx_1,\dots,\bx_4}\rme^{\frac{\rmi}{\hbar}
    S(\bx_1,\dots,\bx_4)}\int_0^t d t' \Theta(t-t'-t_{\rm E}) 
  \delta(t'-t_{\rm E})\\&= -2\tau^2 \Theta(t-2t_{\rm E})}
\end{eqnarray*}
in agreement with the result of the semiclassical analysis.

\subsection{Higher Orders of Perturbation Theory}
\label{sec:higher_orders}

What happens at higher orders in perturbation theory in the parameter
$s^{-1}$? Before turning to the problem in full, it is instructive to
have a look at the zero mode approximation to the model. The action of
the zero mode configuration --- formally obtained by setting
$T(\bx)\equiv T={\,\rm const.}$ --- is given by $S_0[Q]=
\rmi s^+{\rm tr}(\sigma_3^{\rm ar}Q)/4$, where we have used
the standard~\cite{a-Efet83} notation $Q\equiv T^{-1}\Lambda T$.
Parameterizing the matrix $T=1+W$ as in (\ref{woff}), an expansion in
the generators $B$ obtains the expression
\begin{equation}
  \label{szero}
  S_0[B]=\sum_{n=1}^\infty S_0^{(2n)}[B],\qquad
  S_0^{(2n)}[B]=-\rmi s^+{\rm tr}(-B^\dagger B)^n.
\end{equation}
It is known~\cite{a-Efet83} that, term by term in an expansion in
$s^{-1}$, the zero mode functional reproduces the RMT approximation to
the correlation function $R_2(s)$.  Second, there exists a
far--reaching structural connection between the perturbative expansion
of the zero mode theory on the one hand and the Gutzwiller double sum
on the other. (In fact, the correspondence Gutzwiller sum
$\leftrightarrow$ zero dimensional $\sigma$--model $\leftrightarrow$
RMT played a pivotal role in the proof that the semiclassical
expansion coincides with the RMT result.~\cite{a-MuHeBrHaAl04})

More specifically, to each term contributing to the Wick contraction
of
\begin{equation}\label{diagramzero}
  \langle (S_0^{(4)}[B])^{m_2}\,
  (S_0^{(6)}[B])^{m_3}\dots \rangle_0
\end{equation}
there corresponds precisely one semiclassical orbit/partner orbit pair
(or `diagram'). By power counting, this diagram contributes to the
correlation function at order $s^{-2-\sum_n m_n(n-1)}$. For every
value of $n=2,3,\dots$, it contains $m_n$ encounter regions where $n$
orbit segments meet and $\sum_n n m_n$ inter-encounter orbit
stretches. The topology of the diagram is fixed by the way in which
the $B$ matrices are contracted. (For example, the first of the
diagrams shown in figure \ref{tau_3_terms} corresponds to the
contraction $(1-3,2-6,4-8,5-7)$ of ${\rm tr}(B^\dagger B B^\dagger B)
\,{\rm tr}(B^\dagger B B^\dagger B)$, the second diagram to the
contraction $(1-4,2-5,3-6)$ of ${\rm tr}(B^\dagger B B^\dagger B
B^\dagger B)$, etc.)  Importantly, the minimum time required for the
buildup of a diagram (i.e., the time required to traverse the $\sum_n
m_n$ encounter regions) is given by $t_{\rm E}\times \sum_n n m_n$.

Turning back to the full problem, let us consider the analog of the
zero dimensional expression (\ref{diagramzero}), 
\begin{equation}\label{diagramfull}
  \langle (S^{(4)}[B])^{m_2}\,(S^{(6)}[B])^{m_3}\dots \rangle,
\end{equation}
where $S^{(2n)}$ is given by (\ref{s2n}) and the average is over the
full quadratic action~(\ref{actioncl}). It is natural to expect that
the unique correspondence between Wick contractions and semiclassical
diagrams carries over to the full model. If so, individual
contractions should vanish/reduce to the universal RMT result for
times shorter/much larger than $t<t_{\rm E}\times \sum_n n m_n$.  In
section~\ref{sec:pert-expans} this correspondence was exemplified for
the simplest non--trivial example, the Sieber--Richter diagram
$\langle S^{(4)}[B]\rangle$.

\begin{figure}[h]
  \centering
  \epsfig{file=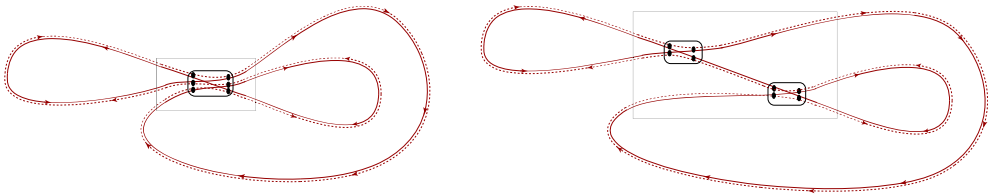, width=\textwidth}
    \caption{\label{S8S6_ambiguity}
    Two representatives of the `clover leaf' diagram class contributing
    to the form factor at ${\cal O}(\tau^3)$. Discussion, see text.}
\end{figure}

Perhaps unexpectedly, the straightforward one--to--one correspondence
outlined above does not pertain to higher orders in perturbation
theory. To anticipate our main findings, it turns out that at order
$(s^{-4}\leftrightarrow \tau^3)$ in the series expansion, propagators
of short duration $P_{t<t_{\rm E}}$ --- absent in the
$(\tau^2\leftrightarrow s^{-3})$ term considered above --- begin to
play a role. This implies that individual contractions may relate to
more than one semiclassical diagram class. Nonetheless, integration
over all time parameters obtains a universal result.

By way of example, let us consider the $(1-3,2-6,4-8,5-7)$ contraction
of $\langle{\rm tr}(B^\dagger B B^\dagger B) {\rm tr}(B^\dagger B
B^\dagger B)\rangle$. For generic values ($t_i\sim t_{\rm H}\gg t_{\rm
  E}$) of the time arguments carried by the four resulting propagators
the contraction corresponds to the orbit pair shown in
figure \ref{tau_3_terms} left.  However, the integration over times
$t_i$ also extends over exceptional values where one of the two
propagators connecting the two encounter regions ($(2-6)$ or $(4-8)$)
is of short duration $<t_{\rm E}$. Such a short time propagator
connects two \emph{distinct\/} vertices.\footnote{While, in principle,
  the theory also permits the formation of short time propagators
  connecting two phase space points of a single vertex, these
  contributions are practically negligible: imagine a propagator
  $P_t(\bx,\bx')$ returning after a short time to its point of
  departure ($|\bx-\bx'|\sim \hbar^\frac{1}{2}$). Since $t$ is much
  shorter than the Ehrenfest time, all other propagators departing
  from the concerned Hikami box will essentially follow the trajectory
  traced out by the return propagator, and, after a time $t$, also
  return to the departure region. In semiclassical language, we are
  dealing with an orbit that traverses a loop structure in phase space
  repeatedly.  It is known, however, that for large time scales, the
  probability to find repetitive orbits is exponentially small (in the
  parameter $\exp(-\lambda t)$), i.e.~short self--retracing
  contractions are negligible.}  This results in a structure as shown
in figure \ref{S8S6_ambiguity} right, where the two clusters of dots
indicate the eight phase space arguments of the $B$--fields, the
straight line--pair represents the short propagator, and the box
indicates that all phase space points lie in a \emph{single\/} encounter
region.  Evidently, this structure corresponds to a pair of orbits
visiting a single encounter region twice. Diagrams of this structure
are canonically obtained by contraction of a `Hikami hexagon' ${\rm
  tr}(B^\dagger B B^\dagger B B^\dagger B)$, as indicated in
figure \ref{S8S6_ambiguity} left.  Fortunately, the absence of a unique
assignment to semiclassical orbit families, does not significantly
complicate the actual computation of the diagrams: closer inspection
shows that taking the Liouville operators involved in the definition
of the Hikami boxes into account and integrating by parts, we again
obtain the universal zero--mode result.

Summarizing, we have seen that at next to leading order in
perturbation theory short time propagators begin to play a role. While
this complication prevents the assignment of Wick contractions to
orbit pairs of definite topology, the results obtained after
integration over all temporal configurations remain universal (agree
with the RMT prediction). We trust that the structures discussed above
are exemplary for the behaviour of the ballistic $\sigma$--model at
arbitrary orders of perturbation theory, i.e.~that after integration
over all intermediate times, each contraction contributing to
(\ref{diagramfull}) produces the universal result otherwise obtained
by its zero dimensional analog equation (\ref{diagramzero}).

\section{Conclusions and Outlook}\label{sec:discussion}

In this paper, we have applied field theoretical methods to explore
quantum interference corrections to the spectral form factor of
individual chaotic systems. We have seen that the formation of the
latter essentially relies on the fact that the ballistic
$\sigma$--model --- a field theory defined in classical phase space
--- is not capable of resolving structures on scales smaller than the
Planck cell. This quantum uncertainty is an intrinsic feature of the
model (viz.~through the fact that the field degrees of freedom are
multiplied by Moyal rather than by conventional products) and need not
be added by hand as was done in previous approaches. In a manner that
largely parallels the results of recent semiclassical analyses, the
interplay of this uncertainty with the instabilities of the underlying
classical chaotic dynamics leads to the formation of universal quantum
interference corrections to the spectral form factor.

The analysis above is perturbative in nature and, thus, limited to
energy scales larger than the single particle level spacing. To
advance into the perturbatively inaccessible regime $\omega<\Delta$
(i.e.~to prove the universality hypothesis in full) one would need to
understand how the conspiracy of quantum uncertainty and classical
instabilities damps out fluctuations inhomogeneous in phase space at
time scales larger than the Ehrenfest time. The identification of a
concrete mechanism effecting this reduction remains an open issue.

\ack

Countless discussions with F.~Haake on all issues relating to the
semiclassical understanding of the problem are gratefully
acknowledged. We have also benefited from discussions with O.~Agam.
Work supported by SFB/TR12 of the Deutsche Forschungsgemeinschaft and
GIF grant 709.

\appendix

\section{Regularization}\label{sec:regularization}

Throughout this appendix we use phase space coordinates
$\bx=(E,t,\by)$, where $E$ is the energy, $t$ a time--like coordinate
conjugate to energy and parameterizing the Hamiltonian flow through
$\bx$, and $\by$ a $(2f-2)$--component vector of energy shell
coordinates transverse to $t$.

\subsection{In--Shell Regularization}

As discussed in the text, the quadratic action of the model is
controlled by the commutator $[H,\;]$ or, upon Wigner transformation,
the series of operators $[H,\;]\mapsto\rmi\hbar\{H,\;\}+ \sum_{n=1}^\infty
\hbar^{2n+1} D^{(2n+1)}(\partial_\bx)$, where
$D^{(2n+1)}(\partial_\bx)$ is an operator of $(2n+1)$--th order in the
phase space derivatives $\{\partial_{x_i}\}$. When acting in a space
of functions smooth on scales $\hbar$, terms beyond the leading term
(the Poisson bracket) are inessential and the quantum dynamics
collapses to its semiclassical limit. Na\"ively, one might hope that to
achieve this reduction it suffices to choose the \emph{initial\/}
distributions in phase space sufficiently smooth. However, what
complicates the problem is that the generator of classical evolution
$\{H,\;\}$ by itself leads to the dynamical buildup of singularities,
no matter how smooth the initial distribution was. The point is that,
due to the global hyperbolicity of the dynamics, we may locally
identify truly expanding and contracting coordinate directions.
Focusing attention on the latter, and linearizing the flow around a
given reference trajectory, the equations of motion controlling the
evolution of a phase space distribution $\rho$ assume the form $\dot
\rho = \{H,\rho\}=\lambda s\partial_s\rho + \dots$, where $s$ is the
coordinate that contracts strongest, $\lambda$ the corresponding
Lyapunov exponent, and the ellipses indicate derivatives in other
coordinate directions. After a time $t\sim \lambda^{-1}\ln(\delta
x_0/\hbar)$, where $\delta x_0$ denotes the characteristic initial
extension of the distribution, structures in the $s$--direction
fluctuating on scales $\lesssim\hbar$ will have formed implying that
derivatives $D^{(2n+1)}$ acting in $s$--direction can no longer be
neglected. One way to remove this complication~\cite{a-ZuPa94} is to
add to the generator of classical time evolution an elliptic operator
$\sim D\partial_s^2$, where $D$ is constant. Indeed, it is
straightforward to show (by dimensional analysis or by explicit
calculation) that for the regularized operator $\lambda s\partial_s +
D \partial^2_s$ the initial contraction halts at a characteristic
scale $s\sim (D/\lambda)^{\frac{1}{2}}$. Choosing $D\sim \hbar^2$, it
is guaranteed that the distribution will not build up structure on
scales $<\hbar$, i.e.~that the quantum corrections to classical
dynamics are negligible.  This motivates the addition of the
regulatory contribution (\ref{actionreg}) to the action.

\subsection{Off--Shell Regularization}

In the main text (cf., e.g., equation (\ref{moyal2nB})), we have assumed
that (a) all fields are defined on a shell of constant energy $E_0$
and (b) the theory is local in the conjugate `time' coordinate $t$. To
understand the meaning of this reduction, we need to recall the
original definition of the field degrees of freedoms, $Q(\bx)$, of the
ballistic $\sigma$--model. According to Ref.~\cite{a-AnSiAgAl96},
$Q=T^{-1} \tilde \Lambda T$, where $\tilde \Lambda \equiv
\delta_{E_{\rm av}}(E_0-H)\otimes\Lambda $ and the `delta function'
\begin{equation*}
\delta_{E_{\rm av}}(E_0-H)\equiv\frac{1}{\pi E_{\rm av}}
\sqrt{1-\Bigl(\frac{E_0-H}{2E_{\rm av}}\Bigr)^2}
\end{equation*}
projects on an energy window of width $E_{\rm av}$. (In the original
reference, $E_{\rm av}$ was identified with the energy window over
which the two--point correlation function (\ref{r2}) is averaged,
hence the subscript `av'.) Integrating an action functional of these
field degrees of freedom over \emph{all\/} of phase space and absorbing
$\delta_{E_{\rm av}}$ into the integration measure, $d\bx \mapsto
d\bx\times \delta_{E_{\rm av}}(E_0-H)\propto (d\bx)$, we obtain the
`energy shell' measure used in the text. To understand the energy
dependence of the fields themselves, we write $T(\bx)=1+W(\bx)$, where
$W(\bx)$ anti--commutes with $\Lambda$ and, therefore, commutes with
the function $\delta_{\rm E_{\rm av}}(E_0-H)$. (Recall that all
products are Moyal products, i.e.~functions in phase space do not
necessarily commute with each other.) Evaluating the latter condition,
we obtain $[W, \delta_{E_{\rm av}}(E_0-H)](\bx) = \rmi\hbar
\delta'_{E_{\rm av}}(E_0-E) \partial_t W + {\cal O}(\hbar^3)$. To
rigorously fulfill this condition, we would need to require
independence of $W$ of the coordinate $t$ along the flow through the
phase space point $\bx$ --- obviously too strong a restriction.
Instead, we will impose the weaker condition of approximate
commutativity, $|| [W,\delta_{E_{\rm av}}(E_0-H)]||^2\ll ||W
\delta_{E_{\rm av}}(E_0-H)||^2$, where the operator norm is defined as
$||A||^2\equiv{\rm tr}(A^\dagger A)=\int\rmd \bx |A(\bx)|^2$. It is
straightforward to check --- by explicit calculation or by dimensional
reasoning --- that the commutator is small in the above sense provided
that $\partial_t W < (E_{\rm av}/\hbar)W$, i.e.~the fields $W$ have to
be smooth in `time direction' on scales $\sim \hbar/E_{\rm
  av}$.\footnote[1]{Using equation (\ref{moyal_nonloc}), it is also
  straightforward to show that field configurations $W(E,\tau)$ which
  rigorously commute with $\delta_{E_{\rm av}}(E_0-H)$ do (a) vanish
  for energies outside a window of width $E_{\rm av}$ around $E_0$ and
  (b) have a bounded Fourier spectrum $|\epsilon|<E_{\rm av}$, where
  $\epsilon$ is Fourier conjugate to the time variable $\tau$.}

The above energy--time duality suggests the following interpretation
of the theory: let us introduce a `stroboscopic' picture of the
particle dynamics wherein time scales smaller than a certain
\emph{classical\/} $t_0$ need not be resolved. (E.g., in a billiard,
$t_0\ll t_{\rm f}$ where $t_{\rm f}$ is the time of flight through the
system, etc.)  All fields are smooth on scales $t_0$. This means that
the width of the averaging window must be (at least) of order $E_{\rm
  av}\lesssim \hbar/t_0$. In the classical limit, we indeed project
onto a sharp `energy shell'. (However, we do not know how to reconcile
the condition of anti--commutativity with $\Lambda$ with the condition
of a \emph{mathematically\/} sharp energy shell proposed
in~\cite{a-EfScTa04}.) Second, we require that the fields $W(E)$ do
not vary significantly over their narrow range $[E_0-E_{\rm av},
E_0+E_{\rm av}]$ of definition (the so--called `mode locking
assumption'~\cite{a-AnSiAgAl96}). This can be achieved by choosing the
second order regulator derivative in $E$ direction as $\sim
(\hbar/t_0)^2\partial^2_E$. We thus integrate over field
configurations that are coarse grained over Planck cells of extension
$(t_0,E_{\rm av}=\hbar/t_0)$. As a result, the integral over the
$(E,t)$--sector of the Moyal products can be carried out and we arrive
at the theory independent of energies and local in time direction
considered in the text.

\section{Perturbation Theory}\label{sec:perturbation_theory}

For completeness, we briefly summarize the contraction
rules~\cite{a-AnSiAgAl96} employed in calculating integrals over
products of $B$ matrices:
\begin{equation}\label{contraction}
\eqalign{\langle\tr(B(\bx) A)\;
\tr(B^\dagger(\bx')A')\rangle_B &=
\frac{\Omega}{t_{\rm H}}P_\omega(\bx,\bx')\,\tr(AA'),\\
\langle\tr(B(\bx) A B^\dagger(\bx') A')\rangle_B &=
\frac{\Omega}{t_{\rm H}}P_\omega(\bx,\bx')\,\tr(A)\,\tr(A'),\\
\langle\tr(B(\bx) A)\;
\tr(B(\bx') A')\rangle_B &=
\frac{\Omega}{t_{\rm H}}P_\omega(\bx,\bar{\bx}')\,\tr(AA^{\prime\tau}),\\
-\langle\tr(B(\bx) A B(\bx') A')\rangle_B &=
\frac{\Omega}{t_{\rm H}}P_\omega(\bx,\bar{\bx}')\,\tr(A A^{\prime\tau}).}
\end{equation}
where $A$ and $A'$ are arbitrary fixed matrices. To compute the
integral over an
arbitrary product of traces of $B$--matrices, one first forms all
possible total pairings
$B$---$B^\dagger$, $B$---$B$, and $B^\dagger$---$B^\dagger$, and then
computes individual pairings by means of (\ref{contraction}). Each
contraction reduces the number of matrices by two. Eventually, one
obtains an expression $\sim (\tr\,\mathbf{1})^n=(2R)^{n\ge 2}$ (where
all contributions with $n>2$ vanish in the replica limit).

\end{document}